\documentclass[12pt]{article}
\pdfoutput=1
\usepackage{geometry,enumerate,amsmath,amssymb}
\usepackage{fullpage}
\usepackage{graphicx}
\usepackage{bm}% bold math
\usepackage[colorlinks=false, urlcolor=blue, linkcolor=red, hidelinks]{hyperref}
\numberwithin{equation}{section}
\newcommand{\be}{\begin{equation}}
\newcommand{\ee}{\end{equation}}
\newcommand{\bea}{\begin{eqnarray}}
\newcommand{\eea}{\end{eqnarray}}
\renewcommand{\epsilon}{\varepsilon}

\begin{document}
\title{Platonic solutions of the discrete Nahm equation}
\author{
  Paul Sutcliffe\\[10pt]
  {\em \normalsize Department of Mathematical Sciences,}\\
{\em \normalsize Durham University, Durham DH1 3LE, United Kingdom.}\\
 {\normalsize Email:  p.m.sutcliffe@durham.ac.uk}
}

\date{March 2026}

\maketitle
\begin{abstract}
  The discrete Nahm equation is an integrable nonlinear difference equation for complex $N\times N$ matrices defined on a one-dimensional lattice, with rank and symmetry boundary conditions at the ends of the lattice. Solutions of this system correspond to $SU(2)$ magnetic monopoles of charge $N$ in hyperbolic space, with the curvature related to the number of lattice points. Here some solutions of the discrete Nahm equation are obtained by imposing platonic symmetries, and the spectral curves of the associated hyperbolic monopoles are calculated directly from these solutions. 

\end{abstract}

\newpage

\section{Introduction}\quad
Atiyah \cite{At,At2} observed that for any positive integer $m$, $SU(2)$ magnetic monopoles of charge $N$ in three-dimensional hyperbolic space with sectional curvature $-1/m^2$ correspond to circle-invariant $SU(2)$ Yang-Mills instantons in four-dimensional Euclidean space with instanton number $mN.$ This correspondence applies in a normalization in which the length of the Higgs field at infinity is equal to $1/2$, but by applying a scaling symmetry this is equivalent to fixing the curvature of hyperbolic space to be $-1$ and setting the length of the Higgs field at infinity equal to $m/2.$ This quantity is often referred to as the monopole mass, hence the hyperbolic monopoles of interest here are known as half-integral mass monopoles.

Braam and Austin \cite{BA} adapted the ADHM construction of instantons \cite{ADHM} to half-integral mass hyperbolic monopoles by imposing circle-invariance, to yield a correspondence between these charge $N$ hyperbolic monopoles and an integrable nonlinear difference equation for complex $N\times N$ matrices defined on a one-dimensional lattice with $m+1$ lattice points, together with boundary conditions on the matrices at the ends of the lattice. In the continuum limit, $m\to\infty$, the difference equation becomes the Nahm equation, which appears in adapting the ADHM construction to the flat space limit of monopoles in three-dimensional Euclidean space \cite{Nahm}. As the Braam-Austin difference equation is an integrable discretization of the Nahm equation it is commonly referred to as the discrete Nahm equation.

The only solution of the discrete Nahm equation presented in the original work of Braam and Austin is for charge $N=1$, when the matrices are simply constant scalars that specify the position of the single hyperbolic monopole. Ward \cite{Ward} obtained the general solution for $N=2$, but did not impose the boundary conditions required for the monopole correspondence. Murray and Singer \cite{MuSi2} identified the appropriate boundary conditions for a special case of Ward's solution in which the $N=2$ monopole has axial symmetry. If $m=1$ then the discrete Nahm equation degenerates to a complex restriction of the quaternionic ADHM construction, allowing a large class of solutions to be obtained \cite{BCS} by embedding the data of the JNR harmonic ansatz for instantons \cite{JNR} within the ADHM formalism. Furthermore, for $m=1$ there is an alternative description of circle-invariant ADHM data that uses a different circle action \cite{MS}, and it is known how to map this data into complex ADHM data \cite{Su}, providing some solutions beyond the JNR class. However, it is not currently understood how to extend this approach to $m>1.$

In this paper the first examples of solutions of the discrete Nahm equation with $m>1$ and $N>2$ are presented, by applying platonic symmetries to simplify the system. The associated monopole spectral curves are obtained directly from these solutions. Some, but not all, of these spectral curves have been obtained previously using methods of algebraic geometry \cite{NoRo}. Section 2 provides a brief review of the discrete Nahm equation. Section 3 describes the imposition of platonic symmetry and presents some examples, including the lowest charge solutions with tetrahedral, octahedral and icosahedral symmetry, with $N=3,4,7,$ respectively.

\section{The discrete Nahm equation}\quad
Consider a finite one-dimensional lattice, with the lattice points labelled by the integers $0,1,...,m,$ where $m$ is taken to be an odd integer to simplify the presentation. Colour the even and odd lattice points black and white respectively, and attach the $N\times N$ complex matrices 
$B_{2\ell}$ and $W_{2\ell+1}$ to the black and white lattice points.
The discrete Nahm equation \cite{BA} is the following set of nonlinear matrix difference equations
\bea
B_{2\ell+2}=W_{2\ell+1}^{-1}B_{2\ell}W_{2\ell+1}, \qquad\quad \ \ 
\qquad &&\mbox{ for }\ \ell=0,1,..,(m-3)/2,\label{Beqn}\\
W_{2\ell+1}W_{2\ell+1}^\dagger=W_{2\ell-1}^\dagger W_{2\ell-1} +[B_{2\ell}^\dagger,B_{2\ell}], \qquad &&\mbox{ for }\ \ell=0,1,..,(m-1)/2,\label{Weqn}
\eea
where $W_{-1}=W_1^t$  and $^\dagger$ denotes the hermitian conjugate. The boundary condition at the first lattice point is
$B_0=B_0^t$ and the boundary condition at the last lattice point is that
$W_m$ has rank one. Note that there is some freedom in obtaining $W_{2\ell+1}$ from the quantity $W_{2\ell+1}W_{2\ell+1}^\dagger$, so Cholesky decomposition will be used to fix $W_{2\ell+1}$ to be a lower triangular matrix.

The discrete Nahm equation  is a discrete integrable system with a spectral curve that is a biholomorphic algebraic curve in $\mathbb{CP}^1\times\mathbb{CP}^1$ of bidegree $(N,N)$, given by \cite{MuSi2}
\be
\mbox{det}\bigg(\eta\zeta B_{2\ell}^\dagger+\zeta-\eta(B_{2\ell}^\dagger B_{2\ell}+W_{2\ell-1}^\dagger W_{2\ell-1})-B_{2\ell}\bigg)=0.
\label{sc}
\ee
This curve is independent of $\ell$ and encodes the conserved quantities for the evolution along the lattice. The same spectral curve is associated with the corresponding hyperbolic monopole fields via a mini-twistor description of geodesics along which a particular scattering equation has decaying solutions \cite{At}.
Rotations act on the spectral curve coordinates as M\"obius transformations
\be
(\eta,\zeta)\mapsto
\bigg(
\frac{\alpha \eta+\beta}{-\bar\beta \eta+\bar\alpha},
\frac{\alpha \zeta+\beta}{-\bar\beta \zeta+\bar\alpha}\bigg),
\label{rot}
\ee
where
\be
g=\begin{pmatrix}\alpha & \beta\\ -\bar\beta & \bar\alpha \end{pmatrix}\in SU(2).\label{SU2}
\ee

To identify the continuum limit, $m\to\infty$, set the lattice spacing to be $\frac{1}{2m}$ and replace the integer lattice ${0,1,...,m},$ by the variable $s\in[0,\frac{1}{2}].$ Introduce the triplet of antihermitian $N\times N$ matrices, $T_1(s),T_2(s),T_3(s),$ and write
\be
B_{2\ell}=iT_2(s)-T_1(s),\qquad W_{2\ell+1}=m+iT_3(s+\frac{1}{2m}).
\ee
Taking the continuum limit, $m\to\infty$, of (\ref{Beqn}) and (\ref{Weqn}) yields the Nahm equation
\be
\frac{dT_i}{ds}=\sum_{j=1}^3\sum_{k=1}^3 \varepsilon_{ijk}T_jT_k,
\ee
where $\varepsilon_{ijk}$ is the totally antisymmetric tensor. This is consistent with the fact that the discrete Nahm equation describes monopoles in hyperbolic space with curvature $-1/m^2$ and the Nahm equation describes monopoles in the flat space Euclidean limit.

\section{Platonic solutions}\quad
Take $G\subset SO(3)$ to be one of the platonic symmetry groups, namely, the tetrahedral ($G=T$), octahedral ($G=O$), or icosahedral ($G=Y$) group. A triplet $(Y_1,Y_2,Y_3)$ of real symmetric $N\times N$ matrices is $G$-symmetric if for each element ${\cal O}\in G$ there exists a matrix $F_{\cal O}\in SO(N)$ such that
\be
\sum_{j=1}^3{\cal O}_{ij}Y_j=F_{\cal O}Y_iF_{\cal O}^{-1}.
\label{Gsymm}
\ee
Methods have been introduced \cite{HMM} and developed \cite{HS1} to obtain such $G$-symmetric triplets of matrices from invariant homogeneous polynomials over $\mathbb{CP}^1$. Furthermore, code is publicly available to automatically calculate these matrices given the invariant polynomials as the input data \cite{DH}. These methods were pioneered to construct platonic solutions of the Nahm equation, where the requirement is to obtain a triplet of antihermitian matrices, but it is easy to adapt the reality structure to the case of real symmetric matrices \cite{Su2}. In this section, a procedure is introduced to obtain platonic solutions of the discrete Nahm equation from the data of a $G$-symmetric triplet of matrices. Some illustrative examples will be presented below.

Given a $G$-symmetric triplet, $(Y_1,Y_2,Y_3),$ that will contain at least one free parameter, use this triplet to define the initial data for the discrete Nahm evolution via the formulae
\bea
B_0&=&(1-Y_3)^{-\frac{1}{2}}(Y_1+iY_2)(1-Y_3)^{-\frac{1}{2}},\label{B0}\\
W_1W_1^\dagger&=&(1-Y_3)^{-\frac{1}{2}}\big(1+Y_3-(Y_1+iY_2)(1-Y_3)^{-1}(Y_1-iY_2)\big)(1-Y_3)^{-\frac{1}{2}}.\label{W1}
\eea
To show that this yields a $G$-symmetric spectral curve, first take (\ref{sc}) with $\ell=0$ to give
\bea
0
&=&\mbox{det}\bigg(\eta\zeta B_0^\dagger+\zeta-\eta(B_0^\dagger B_{0}+W_{-1}^\dagger W_{-1})-B_{0}\bigg)\nonumber\\
&=&\mbox{det}\bigg(\eta\zeta \overline{B_0}+\zeta-\eta(\overline{B_0} B_{0}+\overline{W_1 W_{1}^\dagger})-B_{0}\bigg)\\
&=&\mbox{det}\bigg((1-Y_3)^{-1}\bigg)\,\mbox{det}\bigg(\eta\zeta(Y_1-iY_2)+\zeta(1-Y_3)-\eta(1+Y_3)-(Y_1+iY_2)\bigg).\nonumber
\eea
For each ${\cal O}\in G$, the associated M\"obius transformation (\ref{rot}), with $g\in SU(2)$ given by (\ref{SU2}), is obtained from the double cover formula
\be
   {\cal O}_{ij}=\frac{1}{2}\mbox{Tr}(g\overline{\sigma}_ig^{-1}\overline{\sigma}_j),\label{dcover}
   \ee
   where $\sigma_i$ denote the Pauli matrices.
   On applying the rotation (\ref{rot})
   \bea
    &&\mbox{det}\bigg(\eta\zeta(Y_1-iY_2)+\zeta(1-Y_3)-\eta(1+Y_3)-(Y_1+iY_2)\bigg)\mapsto \nonumber\\
   &&\frac{
     \mbox{det}\bigg(F_{\cal O }\big(
     \eta\zeta(Y_1-iY_2)+\zeta(1-Y_3)-\eta(1+Y_3)-(Y_1+iY_2)\big)F_{\cal O}^{-1}
     \bigg)
}{(-\bar\beta \eta+\bar\alpha)^N(-\bar\beta \zeta+\bar\alpha)^N},
   \eea
   where (\ref{Gsymm}) and (\ref{dcover}) have been used to obtain this expression by exchanging the $\alpha$ and $\beta$ dependence inside the determinant for conjugation of the $Y_i$ by $F_{\cal O}$. This proves that the curve is $G$-symmetric. Finally, the parameters in the $G$-symmetric triplet $(Y_1,Y_2,Y_3)$ are determined by evolving the initial data, (\ref{B0}) and (\ref{W1}), to obtain $W_m$ and imposing the condition that this has rank one. This provides the required platonic solution of the discrete Nahm equation and the associated spectral curve.
   
   The remainder of this section provides some examples of applying this procedure. As mentioned earlier, the lowest charge solutions with platonic symmetry groups $G=T,O,Y$ are $N=3,4,7.$ This can be established using a correspondence between hyperbolic monopoles and rational maps between Riemann spheres \cite{JN}, to convert the question of the existence of a $G$-symmetric hyperbolic monopole with charge $N$ to that of a $G$-symmetric rational map of degree $N$. This can be determined using a little representation theory \cite{HMS}, as follows. Let $\underline{N+1}$ denote the $(N+1)$-dimensional irreducible representation of $SU(2)$, and $\underline{N+1}|_{G^*}$ its restriction to the binary platonic subgroup $G^*\subset SU(2).$ In general $\underline{N+1}|_{G^*}$ will be a reducible representation of $G^*$, and the existence of a symmetric rational map requires that it contains a 2-dimensional irreducible representation of $G^*$ or at least two 1-dimensional irreducible representations. Applying this criterion reveals the above charges as the lowest possible for the given platonic symmetries. The same analysis confirms the existence of a charge $N=5$ hyperbolic monopole with octahedral symmetry. Solutions of the discrete Nahm equation corresponding to all four of these platonic monopoles, with charges $N=3,4,5,7,$ will be presented below. The representation analysis shows that there are no platonic solutions with charge $N=6$.

\subsection{$N=3$ tetrahedral solutions}\quad
In a basis in which $Y_3$ is diagonal, a symmetric triplet with $N=3$ and $G=T$ is given by
\be
Y_1+iY_2=\frac{d}{\sqrt{2}}
\left(\begin{array}{ccc}
0 & 0 & 1-i  
\\
 0 & 0 & 1+i  
\\
1 -i  & 1+i  & 0 
\end{array}\right),
\qquad
Y_3=d
\left(\begin{array}{ccc}
1 & 0 & 0 
\\
 0 & -1 & 0 
\\
 0 & 0 & 0 
\end{array}\right),
\label{Y3}
\ee
where $d$ is a real parameter in the interval $(-1,1),$ to ensure that $1-Y_3$ is a positive definite matrix.
Substituting these matrices into (\ref{B0}) gives
\be
B_0=\frac{d}{\sqrt{2(1-d^2)}}
  \left(\begin{array}{ccc}
0 & 0 & \left(1-{i}\right) \sqrt{1+d} 
\\
 0 & 0 & \left(1+{i}\right) \sqrt{1-d} 
\\
 \left(1-{i}\right) \sqrt{1+d} & \left(1+{i}\right) \sqrt{1-d} & 0 
  \end{array}\right).
  \ee
  Substituting (\ref{Y3}) into (\ref{W1}) and performing Cholesky decomposition produces
    \be
  W_1=
    \left(\begin{array}{ccc}  
      \sqrt{\frac{1+d-d^2}{1-d}} & 0 & 0\\
      \frac{-id^2}{\sqrt{(1+d-d^2)(1+d)}} & \sqrt{\frac{1-3d^2}{(1+d-d^2)(1+d)}} & 0\\
        0 & 0 & \sqrt{\frac{1-3d^2}{1-d^2}}
 \end{array}\right).\label{W1T}
    \ee
The spectral curve is invariant under the
generators of the tetrahedral group
\be
(\eta,\zeta)\mapsto(-\eta,-\zeta),\qquad \qquad
(\eta,\zeta)\mapsto\bigg(\frac{\eta-i}{\eta+i},\frac{\zeta-i}{\zeta+i}\bigg),
\label{tetrotations}
\ee
and takes the form
\be
(\eta-\zeta)^3+ic_T(\eta+\zeta)(\eta^2\zeta^2-1)=0,
\ee
where the coefficient is
\be
c_T=\frac{2d^3}{1-d^2}.
\label{ctet}
\ee
Note that changing the sign of $d$ is equivalent to a rotation, so without loss of generality $d$ may be taken to be positive.
Calculating the determinant of (\ref{W1T}) gives the result
\be
\mbox{det}(W_1)=\frac{1-3d^2}{1-d^2},
\ee
which vanishes if $d$ takes the value $d_1=\frac{1}{\sqrt{3}},$ so that $W_1$ has rank one, providing a solution to the discrete Nahm equation with $m=1.$ Substituting $d=d_1$ into (\ref{ctet}) yields the spectral curve coefficient
$c_T=\frac{1}{\sqrt{3}}$, for this solution of the system with $m=1.$ This reproduces the earlier result obtained for this spectral curve using methods of algebraic geometry \cite{NoRo}.

$W_1$ has full rank for general $d<d_1$, so the evolution may be continued along the lattice using (\ref{Beqn}) and (\ref{Weqn}) to obtain
\be
B_2=  \frac{d}{\sqrt{2(1+d-d^2)}}
\left(\begin{array}{ccc}
  0 & 0 & (1-i)
  \sqrt{\frac
    {1-3d^2}{1-d^2}}\\
  0 & 0 & (1+i)
  \sqrt{\frac
    {1+d}{1-d}}\\
  (1-i)\sqrt{\frac
    {(1+2d+d^2-2d^3)^2}{(1-d^2)(1-3d^2)}} &
  (1+i)\sqrt{\frac
    {1-d}{1+d}} & 0
\end{array}\right),
\ee
and
\be
W_3=
\left(\begin{array}{ccc}
  \sqrt{\frac{p_3}{p_1p_2(1-d^2)}} & 0 & 0\\
  id^2p_4\sqrt{\frac{1+d}{(1-d)p_1p_3}} & \sqrt{\frac{(1+d^2)p_1p_5}{p_3}} & 0\\
0 & 0 & \sqrt{\frac{(1+d^2)p_5}{(1-d^2)p_2}}
\end{array}\right),\label{W3T}
\ee 
where the following polynomials have been introduced for notational convenience
\be
\begin{split}
  &p_1=1+d-d^2, \ p_2=1-3d^2, \ p_3=4 d^{8}-4 d^{7}-16 d^{6}+9 d^{5}+9 d^{4}-8 d^{3}-2 d^{2}+3 d +1,\nonumber \\
  & p_4=2 d^{3}-5 d^{2}-2 d +3, \ p_5=4 d^{4}-7 d^{2}+1.
\end{split}
\ee

Calculating the determinant of (\ref{W3T}) gives
\be
\mbox{det}(W_{3})=\frac{(1+d^2)(4 d^{4}-7 d^{2}+1)}{(1-d^2)(1-3d^2)},
\ee
which vanishes if $d$ takes the value $d_3=(\sqrt{11}-\sqrt{3})/4$. For this value, $W_3$ has rank one, providing a solution to the discrete Nahm equation with $m=3.$ Substituting $d=d_3$ into (\ref{ctet}) gives the value of the associated spectral curve coefficient $c_T=2\sqrt{3}-\sqrt{11}$, which again agrees with the result from algebraic geometry \cite{NoRo}.
For $d<d_3$ the matrix $W_3$ has full rank and the evolution may be continued along the lattice using (\ref{Beqn}) and (\ref{Weqn}) to determine the critical value $d_5$ for the vanishing of $\mbox{det}(W_{5})$ and the associated spectral curve coefficient $c_T$ for the $m=5$ solution, and so on. Table 1 presents the values of $c_T$ for odd $m\le 13.$ A scaling argument on the spectral curve coordinates shows that $c_T\to 0$ as $m\to\infty$, due to the way that geodesics in hyperbolic space are defined in terms of mini-twistor variables and their relation to those in Euclidean space \cite{NoRo}.
\begin{table}[h!]
\begin{center}
  \begin{tabular}{|c|c|c|c|c|}
    \hline
    $m$  & $c_T$ & $c_O$ &  $\widetilde c_O$ &$c_Y$ \\
    \hline
    1 & 0.5774 & 0.3333 & 1.0000 & 2.0000\\
    3 & 0.1475 & 0.0718 & 0.2000 & 0.3333\\
    5 & 0.0593 & 0.0243 & 0.0704 & 0.0902\\
    7 & 0.0297 & 0.0105 & 0.0316 & 0.0317\\
    9 & 0.0170 & 0.0053 & 0.0164 & 0.0132\\
    11& 0.0106 & 0.0029 & 0.0094 & 0.0062\\
    13& 0.0071 & 0.0018 & 0.0057 & 0.0032\\
    \hline
  \end{tabular}
  \caption{The spectral curve coefficients, $c_T,c_O,\widetilde c_O,c_Y$, to 4 decimal places, for the odd values of $m$ from 1 to 13.}
\end{center}
\end{table}
  
\subsection{$N=4$ octahedral solutions}\label{sec4O}\quad
A symmetric triplet with $N=4$ and $G=O$ is given by
\be
Y_1+iY_2=\frac{d}{2}
\left(\begin{array}{cccc}
i \sqrt{3} & -\sqrt{3} & i  & -1 
\\
 -\sqrt{3} & -i \sqrt{3} & 1 & i  
\\
 i  & 1 & -i \sqrt{3} & -\sqrt{3} 
\\
 -1 & i  & -\sqrt{3} & i \sqrt{3} 
\end{array}\right),
\qquad
Y_3=d
\left(\begin{array}{cccc}
-1 & 0 & 0 & 0 
\\
 0 & -1 & 0 & 0 
\\
 0 & 0 & 1 & 0 
\\
 0 & 0 & 0 & 1 
\end{array}\right),
\label{Y4}
\ee
where the restriction $d\in(0,1)$ guarantees that $1-Y_3$ is positive definite. Applying (\ref{B0}) and (\ref{W1}), this triplet generates the matrices
\be
B_0=\frac{d}{2}
\left(\begin{array}{cccc}
\frac{ i   \sqrt{3}}{1+d} & \frac{ -\sqrt{3}}{1+d} & \frac{ i }{\sqrt{1-d^2}} & \frac{-1}{\sqrt{1-d^2}} 
\\
 \frac{- \sqrt{3}}{1+d} & \frac{-i\sqrt{3}}{1+d} & \frac{1}{\sqrt{1-d^2}} & \frac{ i }{\sqrt{1-d^2}} 
\\
 \frac{ i }{\sqrt{1-d^2}} & \frac{1}{\sqrt{1-d^2}} & \frac{ -i \sqrt{3}}{1-d} & \frac{- \sqrt{3}}{1-d} 
\\
 \frac{-1}{\sqrt{1-d^2}} & \frac{ i}{\sqrt{1-d^2}} & \frac{-\sqrt{3}}{1- d} & \frac{{i} \sqrt{3}}{1- d}, 
\end{array}\right),
\ee
\be
W_1=
\left(\begin{array}{cccc}
  \frac{1}{1+d}\sqrt{\frac{(1-2d)q_+}{1-d}} & 0 & 0 & 0\\
  \frac{-id^2}{1+d}\sqrt{\frac{1-2d}{(1-d)q_+}} &
  \sqrt{\frac{1-4d^2}{(1+d)q_+}} & 0 & 0\\
    0 & 0 & \frac{1}{1-d}\sqrt{\frac{(1+2d)q_-}{1+d}}& 0\\
0 & 0& \frac{id^2}{1-d}\sqrt{\frac{1+2d}{(1+d)q_-}} & \sqrt{\frac{1-4d^2}{(1-d)q_-}}   
\end{array}\right),
\label{W1O}
\ee
where $q_\pm=1-d^2\pm d.$ Defining
\be
c_O=\frac{3d^4}{(1-d^2)^2},
\label{coct}
\ee
the octahedrally symmetric spectral curve is
\be
(\eta-\zeta)^4+c_{O}(\eta^4\zeta^4+6\eta^2\zeta^2+4\eta\zeta(\eta^2+\zeta^2)+1)=0,
\ee
invariant under the generators of the octahedral group
\be
(\eta,\zeta)\mapsto(i\eta,i\zeta),\qquad \qquad
(\eta,\zeta)\mapsto\bigg(\frac{\eta-i}{\eta+i},\frac{\zeta-i}{\zeta+i}\bigg).
\label{octrotations}
\ee
The determinant of (\ref{W1O}) is
\be
\mbox{det}(W_1)=\frac{(1-4d^2)^\frac{3}{2}}{(1-d^2)^2},
\ee
with $W_1$ having rank one if $d$ is equal to $d_1=\frac{1}{2}$, so that $c_O=\frac{1}{3}.$ This provides the $m=1$ solution of the discrete Nahm equation. If $d<d_1$ the system is evolved along the lattice to produce matrices $B_2$ and $W_3$ with
\be
\mbox{det}(W_3)=\frac{(1+2d^2)(1+2d-2d^2)^\frac{3}{2}(1-2d-2d^2)^\frac{3}{2}}
{(1-d^2)^2(1-4d^2)},
\ee
which vanishes for $d$ equal to $d_3=(\sqrt{3}-1)/2$. The matrices for this $m=3$ solution are
\be
B_2=
\left(\begin{array}{cccc}
  \frac{i(5\sqrt{3}-3)}{22} &
  \frac{3\sqrt{2}(4-3\sqrt{3})}{22} &
  \frac{i\sqrt{66 \sqrt{3}+396}}{66}
    & \frac{-\sqrt{22+44 \sqrt{3}}}{22} 
\\ \\
\frac{-\sqrt{2}(1+2\sqrt{3})}{22}
& \frac{i(3-5\sqrt{3})}{22} &
\frac{\sqrt{198-88 \sqrt{3}}}{22} &
\frac{i\sqrt{-264+198 \sqrt{3}}}{22} 
\\ \\
\frac{i \sqrt{-15774 \sqrt{3}+27324}}{22} &
\frac{\sqrt{2442-1408 \sqrt{3}}}{22}
& \frac{i \left(\sqrt{3}-3\right)}{2} &
\frac{-\sqrt{-6+4 \sqrt{3}}}{2} 
\\ \\
\frac{-\sqrt{-15774+9108 \sqrt{3}}}{22}
& \frac{i \sqrt{-1408+814 \sqrt{3}}}{22} &
\frac{-\sqrt{-18+12 \sqrt{3}}}{2}
& \frac{i (3-\sqrt{3})}{2} 
\end{array}\right),
\ee
and the rank one matrix
\be
W_3=
\left(\begin{array}{cccc}
0 & 0 & 0 & 0 
\\
 0 & 0 & 0 & 0 
\\
 0 & 0 & 2 \sqrt{-15+9 \sqrt{3}} & 0 
\\
 0 & 0 & 2 i \sqrt{-5 \sqrt{3}+9} & 0 
\end{array}\right).
\ee
Substituting $d=d_3$ into (\ref{coct}) gives the coefficient $c_O=7-4\sqrt{3}$ for $m=3.$ The above values of $c_O$ for $m=1$ and $m=3$ agree with those reported in \cite{NoRo}, and numerical values for larger $m$ are presented in Table 1, obtained by continuing the evolution along the lattice to calculate the values of $d$ satisfying the condition $\mbox{det}(W_m)=0.$

\subsection{$N=5$ octahedral solutions}\label{sec5O}\quad
A symmetric triplet with $N=5$ and $G=O$ is given by
\be
Y_1+iY_2=\frac{d}{4}\left(\begin{array}{ccccc}
0 & \sqrt{2} & 0 & -i \sqrt{2} & 0 
\\
 \sqrt{2} & 0 & 2 \sqrt{3} & 0 & \sqrt{2} 
\\
 0 & 2 \sqrt{3} & 0 & i2\sqrt{3} & 0 
\\
 -i \sqrt{2} & 0 & i2\sqrt{3} & 0 & -i\sqrt{2} 
\\
 0 & \sqrt{2} & 0 & -i\sqrt{2} & 0 
\end{array}\right),
\qquad
Y_3=d\left(\begin{array}{ccccc}
-1 & 0 & 0 & 0 & 0 
\\
 0 & 0 & 0 & 0 & 0 
\\
 0 & 0 & 0 & 0 & 0 
\\
 0 & 0 & 0 & 0 & 0 
\\
 0 & 0 & 0 & 0 & 1 
\end{array}\right),
\label{Y5}
\ee
with $d\in(0,1)$. Applying (\ref{B0}) and (\ref{W1}), this triplet generates the matrices
\be
B_0=\frac{d}{4}\left(\begin{array}{ccccc}
0 & \frac{\sqrt{2}}{\sqrt{1+d}} & 0 & \frac{-i \sqrt{2}}{\sqrt{1+d}} & 0 
\\
 \frac{\sqrt{2}}{\sqrt{1+d}} & 0 & 2 \sqrt{3} & 0 & \frac{\sqrt{2}}{\sqrt{1-d}} 
\\
 0 & 2 \sqrt{3} & 0 & i2\sqrt{3} & 0 
\\
 \frac{-i\sqrt{2}}{\sqrt{1+d}} & 0 & i2\sqrt{3} & 0 & \frac{-i \sqrt{2}}{\sqrt{1-d}} 
\\
 0 & \frac{\sqrt{2}}{\sqrt{1-d}} & 0 & \frac{-i\sqrt{2}}{\sqrt{1-d}} & 0 
\end{array}\right),
\ee
\be
W_1=\left(\begin{array}{ccccc}
  \sqrt{\frac{\widetilde q}{4(1+d)}} & 0 & 0 & 0 & 0\\
  0 & \sqrt{\frac{(2-d^2)(2-3d^2)}{4(1-d^2)}} & 0 & 0 & 0\\
  0& 0 & \sqrt{\frac{2-3d^2}{2}} & 0 & 0\\
  0 & -id^2\sqrt{\frac{2-3d^2}{4(1-d^2)(2-d^2)}} & 0 & \sqrt{\frac{2-3d^2}{2-d^2}} & 0\\
  \frac{-d^2}{\sqrt{4(1-d)\widetilde q}} & 0 & 0 & 0 & \sqrt{\frac{2(2-3d^2)}{(1-d)\widetilde q}}
  \end{array}\right),
\label{W1O5}
\ee
where $\widetilde q=4-d^2-4d.$ Defining
\be
\widetilde c_O=\frac{3d^4}{4(1-d^2)}
\label{coct5}
\ee
the octahedrally symmetric spectral curve is
\be
(\eta-\zeta)^5-\widetilde c_{O}(\eta-\zeta)(\eta^4\zeta^4+6\eta^2\zeta^2+4\eta\zeta(\eta^2+\zeta^2)+1)=0,
\ee
invariant under the generators (\ref{octrotations}) of the octahedral group.
The determinant of (\ref{W1O5}) is
\be
\mbox{det}(W_1)=\frac{(2-3d^2)^2}{4(1-d^2)},
\ee
with $W_1$ having rank one if $d$ is equal to $d_1=\sqrt{\frac{2}{3}}$, so that $\widetilde c_O=1.$ This provides the $m=1$ solution of the discrete Nahm equation, with a spectral curve in agreement with that obtained using the JNR approach \cite{BCS} that is applicable for $m=1$. If $d<d_1$ the system is evolved along the lattice to produce matrices $B_2$ and $W_3$ with
\be
\mbox{det}(W_3)=\frac{(2+d^2)(2-5d^2)^2}{4(2-3d^2)(1-d^2)},
\ee
which vanishes for $d$ equal to $d_3=\sqrt{\frac{2}{5}}$. The matrices for this $m=3$ solution are
\be
B_2=
\left(\begin{array}{ccccc}
0 & \frac{\sqrt{15}}{10 \sqrt{9-2\sqrt{10}}} & 0 & \frac{-1}{2\sqrt{9-2 \sqrt{10}}}i & 0 
\\
 \frac{\sqrt{15} \left(3-\sqrt{10}\right)}{6 \sqrt{9-2\sqrt{10}}} & 0 & \frac{3 \sqrt{10}}{20} & 0 & \frac{5\sqrt{6}}{\sqrt{9-2\sqrt{10}}\, \left(20-4 \sqrt{10}\right)} 
\\
 0 & \frac{\sqrt{10}}{4} & 0 & \frac{\sqrt{6}}{4}i & 0 
\\
 \frac{ \left(-3+\sqrt{10}\right)}{2\sqrt{9-2 \sqrt{10}}}i & 0 & \frac{\sqrt{6}}{4}i & 0 & \frac{3 \sqrt{10}}{\sqrt{9-2 \sqrt{10}}\, \left(-20+4 \sqrt{10}\right)}i 
\\
 0 & \frac{\sqrt{6}\, \left(5-\sqrt{10}\right)}{20 \sqrt{9-2\sqrt{10}}} & 0 & \frac{\left(-5+\sqrt{10}\right) \sqrt{10}}{20\sqrt{9-2 \sqrt{10}}}i & 0 
\end{array}\right),
\ee
and the rank one matrix
\be
W_3=\frac{\sqrt{9-2\sqrt{10}}}{205}
\left(\begin{array}{ccccc}
85-13\sqrt{10} & 0 & 0 & 0 & 0
\\
0 & 0 & 0 & 0 & 0
\\
0 & 0 & 0 & 0 & 0
\\
0 & 0 & 0 & 0 & 0
\\
-150-47\sqrt{10} & 0 & 0 & 0 & 0
\end{array}\right).
\ee
Substituting $d=d_3$ into (\ref{coct5}) gives the $m=3$ spectral curve coefficient $\widetilde c_O=\frac{1}{5}$. Continuing the evolution along the lattice for $d<d_3$, and calculating the values of $d$ with $\mbox{det}(W_m)=0,$ produces the numerical values of $\widetilde c_O$ presented in Table 1 for larger values of $m$.

\subsection{$N=7$ icosahedral solutions}\quad
A symmetric triplet with $N=7$ and $G=Y$ is given by
\be
Y_1+iY_2=\frac{d}{2\sqrt{2}}
\left(\begin{array}{ccccccc}
\sqrt{2}\, i  & -\sqrt{2} & 0 & i  & 1 & 0 & 0 
\\
 -\sqrt{2} & -\sqrt{2}\, i  & 0 & 1 & -i  & 0 & 0 
\\
 0 & 0 & 0 & 2 \sqrt{2} & 2 \sqrt{2}\, i  & 0 & 0 
\\
 i  & 1 & 2 \sqrt{2} & 0 & 0 & i  & 1 
\\
 1 & -i  & 2 \sqrt{2}\, i  & 0 & 0 & 1 & -i  
\\
 0 & 0 & 0 & i  & 1 & -\sqrt{2}\, i  & \sqrt{2} 
\\
 0 & 0 & 0 & 1 & -i  & \sqrt{2} & \sqrt{2}\, i  
\end{array}\right),
\ee
with $Y_3$ the diagonal matrix $Y_3=\mbox{diag}(d,d,0,0,0,-d,-d)$, for $d\in(0,1).$
The matrices obtained from (\ref{B0}) and (\ref{W1}) are
\be
B_0=\frac{d}{2\sqrt{2}}
\left(\begin{array}{ccccccc}
\frac{i \sqrt{2}}{1-d} & \frac{-\sqrt{2}}{1-d} & 0 & \frac{i}{\sqrt{1-d}} & \frac{1}{\sqrt{1-d}} & 0 & 0 
\\
 \frac{-\sqrt{2}}{1-d} & \frac{-i \sqrt{2}}{1-d} & 0 & \frac{1}{\sqrt{1-d}} & \frac{-i}{\sqrt{1-d}} & 0 & 0 
\\
 0 & 0 & 0 & 2 \sqrt{2} & 2 \,i \sqrt{2} & 0 & 0 
\\
 \frac{i}{\sqrt{1-d}} & \frac{1}{\sqrt{1-d}} & 2 \sqrt{2} & 0 & 0 & \frac{i}{\sqrt{1+d}} & \frac{1}{\sqrt{1+d}} 
\\
 \frac{1}{\sqrt{1-d}} & \frac{-i}{\sqrt{1-d}} & 2 \,i \sqrt{2} & 0 & 0 & \frac{1}{\sqrt{1+d}} & \frac{-i}{\sqrt{1+d}} 
\\
 0 & 0 & 0 & \frac{i}{\sqrt{1+d}} & \frac{1}{\sqrt{1+d}} & \frac{-i \sqrt{2}}{1+d} & \frac{\sqrt{2}}{1+d} 
\\
 0 & 0 & 0 & \frac{1}{\sqrt{1+d}} & \frac{-i}{\sqrt{1+d}} & \frac{\sqrt{2}}{1+d} & \frac{i \sqrt{2}}{1+d} 
\end{array}\right),
\ee
\be
W_1=
\left(\begin{array}{ccccccc}
  \frac{\sqrt{r_1}}{2(1-d)} & 0& 0& 0& 0& 0& 0\\
  \frac{-id^2(1+d)}{2(1-d)\sqrt{r_1}} & \sqrt{\frac{2r_2r_3}{(1-d)r_1}}& 0& 0& 0& 0& 0\\
  0 & 0 & \sqrt{r_2} & 0& 0& 0& 0\\
  0 & 0 & 0 & \sqrt{\frac{r_2r_4}{2(1-d^2)}}  & 0& 0& 0\\
  0 & 0 & 0 & \frac{-id^2}{\sqrt{2}}\sqrt{\frac{r_2}{(1-d^2)r_4}}  &\sqrt{\frac{2r_2}{r_4}}
  & 0& 0\\
  \frac{-d^2}{2}\sqrt{\frac{1-d}{(1+d)r_1}}
  & \frac{-id^2}{\sqrt{2}}\sqrt{\frac{r_2}{(1+d)r_1r_3}} & 0 & 0 & 0 &
  \frac{1}{1+d}\sqrt{\frac{r_2r_5}{2r_3}} & 0\\
  \frac{id^2}{2}\sqrt{\frac{1-d}{(1+d)r_1}} &
  -\frac{d^2}{\sqrt{2}}\sqrt{\frac{r_2}{(1+d)r_1r_3}} & 0 & 0 & 0 &
      \frac{-id^2}{1+d}\sqrt{\frac{r_2}{2r_3r_5}} &
      \sqrt{\frac{4r_2}{(1+d)r_5}}\\
\end{array}\right),\nonumber
\ee
where the following polynomials have been defined
\be
  r_1=4-7 d^{2}+d^{3}, \ r_2=1-2d^2, \ r_3=2+2d-d^{2}, \
   r_4=2-d^2, \ r_5=4+4 d -d^{2}.
   \ee
The spectral curve takes the icosahedrally symmetric form
\be
(\eta-\zeta)\bigg\{(\eta-\zeta)^6+c_Y\bigg(
i\left(\eta +\zeta \right) \left(\eta^{5} \zeta^{5}+1\right)
+5\eta^{2} \zeta^{2} \left(\eta +\zeta \right)^{2}
+ \eta  \zeta \left(\eta^{4}+\zeta^{4}\right)
\bigg)\bigg\}=0,
\ee
where the coefficient is
\be
c_Y=\frac{4d^6}{(1-d^2)^2}.
\label{cicos}
\ee
This curve is invariant under the generators of the icosahedral group
\be
(\eta,\zeta)\mapsto (\omega^4\eta,\omega^4\zeta),
\qquad\qquad
(\eta,\zeta)\mapsto \bigg(
\frac{(\bar\omega^4-\omega^4)\eta+\omega^7-\bar\omega^9}{(\omega^9-\bar\omega^7)\eta+\omega^4-\bar\omega^4},
\frac{(\bar\omega^4-\omega^4)\zeta+\omega^7-\bar\omega^9}{(\omega^9-\bar\omega^7)\zeta+\omega^4-\bar\omega^4},
\bigg),
\ee
where $\omega=e^{i\pi/10}.$
Using the above matrix
\be
\mbox{det}(W_1)=\frac{(1-2d^2)^3}{(1-d^2)^2},
\ee
with $W_1$ having rank one if $d$ equals $d_1=\frac{1}{\sqrt{2}}$, and (\ref{cicos}) giving $c_Y=2$ for the $m=1$ spectral curve. If $d<d_1$ then continuing the evolution along the lattice yields
\be
\mbox{det}(W_3)=\frac{(1+d^2)(1-3d^2)^3}{(1-2d^2)(1-d^2)^2},
\ee
and $W_3$ has rank one if $d$ equals $d_3=\frac{1}{\sqrt{3}}$,
with (\ref{cicos}) giving $c_Y=\frac{1}{3}$ for the $m=3$ spectral curve coefficient. The expression for $B_2$ for this $m=3$ solution is too hefty to reproduce here, but $W_3$ is manageable, being given by
\be
W_3=\frac{1}{105339}
\left(\begin{array}{ccccccc}
949 \sqrt{28416+13986 \sqrt{3}} & 0 & 0 & 0 & 0 & 0 & 0 
\\
 -i\,73 \sqrt{2779218+1266954 \sqrt{3}} & 0 & 0 & 0 & 0 & 0 & 0 
\\
 0 & 0 & 0 & 0 & 0 & 0 & 0 
\\
 0 & 0 & 0 & 0 & 0 & 0 & 0 
\\
 0 & 0 & 0 & 0 & 0 & 0 & 0 
\\
 -74 \sqrt{3817827-2200731 \sqrt{3}} & 0 & 0 & 0 & 0 & 0 & 0 
\\
 i\, 481 \sqrt{54750-31536 \sqrt{3}} & 0 & 0 & 0 & 0 & 0 & 0 
\end{array}\right).
\ee
Continuing the evolution along the lattice for larger $m$, 
and calculating the values of $d$ at which $\mbox{det}(W_m)=0$, produces the numerical values for $c_Y$ shown in Table 1. The methods of algebraic geometry described in \cite{NoRo} were not applied to the icosahedral case, so values of $c_Y$ are not available for comparison from that approach.

\subsection{$N=4$ tetrahedral solutions}\quad
As an example of a one-parameter family of solutions, consider the following symmetric triplet with $N=4$ and $G=T$, where $a\in(-1,1)$ is the free parameter
\bea
Y_1&=&\frac{d}{2}
\left(\begin{array}{cccc}
\frac{-3 \sqrt{2}\, a}{\sqrt{2 a^{2}+4}} & -\sqrt{3} & 0 & \frac{2 a^{2}-2}{\sqrt{2 a^{2}+4}} 
\\
 -\sqrt{3} & \frac{3 \sqrt{2}\, a}{\sqrt{2 a^{2}+4}} & \frac{2-2 a^{2}}{\sqrt{2 a^{2}+4}} & 0 
\\
 0 & \frac{2-2 a^{2}}{\sqrt{2 a^{2}+4}} & \frac{-3 \sqrt{2}\, a}{\sqrt{2 a^{2}+4}} & -\sqrt{3} 
\\
 \frac{2 a^{2}-2}{\sqrt{2 a^{2}+4}} & 0 & -\sqrt{3} & \frac{3 \sqrt{2}\, a}{\sqrt{2 a^{2}+4}} 
\end{array}\right),\\
Y_2&=&\frac{d}{2}
\left(\begin{array}{cccc}
\sqrt{3} & \frac{3 \sqrt{2}\, a}{\sqrt{2 a^{2}+4}} & \frac{2-2 a^{2}}{\sqrt{2 a^{2}+4}} & 0 
\\
 \frac{3 \sqrt{2}\, a}{\sqrt{2 a^{2}+4}} & -\sqrt{3} & 0 & \frac{2-2 a^{2}}{\sqrt{2 a^{2}+4}} 
\\
 \frac{2-2 a^{2}}{\sqrt{2 a^{2}+4}} & 0 & -\sqrt{3} & \frac{-3 \sqrt{2}\, a}{\sqrt{2 a^{2}+4}} 
\\
 0 & \frac{2-2 a^{2}}{\sqrt{2 a^{2}+4}} & \frac{-3 \sqrt{2}\, a}{\sqrt{2 a^{2}+4}} & \sqrt{3} 
\end{array}\right),\qquad
Y_3=d\sqrt{\frac{2+a^2}{2}}
\left(\begin{array}{cccc}
-1 & 0 & 0 & 0 
\\
 0 & -1 & 0 & 0 
\\
 0 & 0 & 1 & 0 
\\
 0 & 0 & 0 & 1 
\end{array}\right).\nonumber
\eea
The associated tetrahedrally symmetric curve, invariant under the generators (\ref{tetrotations}), is
\be
(\eta-\zeta)^4+
c\big(\eta^{4} \zeta^{4}+6 \eta^{2} \zeta^{2}+4 \eta  \zeta \left(\eta^{2}+\zeta^{2}\right)+1\big)
-i\widetilde c
\left(\eta^{2}-\zeta^{2}\right) \left(\eta^{2} \zeta^{2}-1\right)=0,
\ee
where the coefficients are given by
\be
c=\frac{12 \left(2 a^{2}+1\right) d^{4}}{\left(2-\left(a^{2}+2\right) d^{2}\right)^{2}}, \qquad\qquad
\widetilde c=\frac{24 \sqrt{6}\, a \,d^{3}}{\left(2-\left(a^{2}+2\right) d^{2}\right)^{2}}.
\ee
If $a=0$ then the symmetry is enhanced to octahedral symmetry, and the results of Section \ref{sec4O} are reproduced. The $m=1$ solution, for general $a$, is obtained by requiring that
$W_1$ has rank one, which is achieved by setting $d=\sqrt{2/(8+a^2)}$, producing the $m=1$ spectral curve
coefficients
\be
c=\frac{1+2a^2}{3}, \qquad\qquad
\widetilde c=2a\sqrt{\frac{8+a^2}{3}}.
\ee
As $a\to \pm 1$ the curve becomes a product of four curves for charge one hyperbolic monopoles on the vertices of a tetrahedron at infinity.

\section{Conclusion}\quad
A procedure has been described to obtain symmetric solutions of the discrete Nahm equation and the associated hyperbolic monopole spectral curves, for a set of discrete values of the curvature of hyperbolic space. The method has been illustrated by providing the details of some examples with platonic symmetry. In the simplest examples the symmetric matrices depend on a single parameter $d$, and the value of $d$ required by the boundary conditions is fixed by a vanishing determinant condition at the end of the lattice. As the discrete Nahm equation is an integrable system, it may be the case that the polynomials in $d$, that appear in the determinant expressions as the evolution proceeds along the lattice, have interesting properties that make them worthy of further investigation. Other tractable platonic examples are readily available, together with symmetric examples of families of solutions that are obtained by relaxing the platonic symmetry to a subgroup, such as cyclic or dihedral symmetry. These could be investigated using the same method introduced here.

In the case of maximal curvature, $m=1$, an equivalent description of the discrete Nahm data is known \cite{MS} that uses an alternative circle action to reduce the instanton to a hyperbolic monopole. It is not known how to extend this alternative approach to $m>1$, but there should be another corresponding discrete Nahm equation, and it would be interesting to obtain this. Another possible avenue of research is to extend the results in this paper to gauge groups beyond $SU(2).$

\end{document}